\documentclass[nofootinbib,aps,prc]{revtex4-1} 
\usepackage{amsmath}
\usepackage{amssymb}
\usepackage{amsthm}
\usepackage{amsfonts}
\usepackage{xspace}
\usepackage{graphicx}
\usepackage{graphics}
\usepackage{slashed}
\usepackage{bbold}

\usepackage{hyperref}

\newcommand{\CA}{\mathcal{C}^{(^3 \! S_1-^1 \! P_1)}}
\newcommand{\CB}{\mathcal{C}^{(^1 \! S_0-^3 \! P_0)}_{(\Delta I=0)}}
\newcommand{\CC}{\mathcal{C}^{(^1 \! S_0-^3 \! P_0)}_{(\Delta I=1)}}
\newcommand{\CD}{\mathcal{C}^{(^1 \! S_0-^3 \! P_0)}_{(\Delta I=2)}}
\newcommand{\CE}{\mathcal{C}^{(^3 \! S_1-^3 \! P_1)}}

\newcommand{\eftnopi}{$\text{EFT}{(\slashed{\pi})}$\xspace}
\newcommand{\N}{N}

\newcommand{\LRd}{\overset{\leftrightarrow}{D}}

\newcommand{\wave}[3]{\ensuremath{{}^{#1}\mathrm{#2}_{#3}}\xspace}
\newcommand{\oneS}{\wave{1}{S}{0}}

\newcommand{\threeS}{\wave{3}{S}{1}}

\newcommand{\NC}{\ensuremath{N_c}\xspace}

\newcommand{\calO}{\ensuremath{\mathcal{O}}}
\newcommand{\calI}{\ensuremath{\mathcal{I}}}
\newcommand{\calG}{\ensuremath{\mathcal{G}}}
\newcommand{\calC}{\ensuremath{\mathcal{C}}}
\newcommand{\calA}{\ensuremath{\mathcal{A}}}

\newcommand{\pplus}{\mathbf{p}_+}
\newcommand{\pminus}{\mathbf{p}_-}
\newcommand{\vsig}{\vec{\sigma}}
\newcommand{\vtau}{\vec{\tau}}

\usepackage{ulem}

\renewcommand{\emph}[1]{\textit{#1}}

\begin{document}

\title{
Large-$N_c$ limit reduces the number of independent few-body parity-violating low-energy constants in pionless effective field theory}

\author{Matthias R.~Schindler}
\email{mschindl@mailbox.sc.edu}
\affiliation{Department of Physics and Astronomy,University of South Carolina, Columbia, SC 29208, USA}

\author{Roxanne P.~Springer}
\email{rps@phy.duke.edu}
\affiliation{Department of Physics, Duke University, Durham, NC 27708, USA}

\author{Jared Vanasse}
\email{jjv9@phy.duke.edu}
\affiliation{Department of Physics, Duke University, Durham, NC 27708}
\affiliation{Department of Physics and Astronomy, Ohio University, Athens, OH 45701}

\date{\today}

\begin{abstract}

The symmetries of the Standard Model dictate that for very low energies, where nucleon dynamics can be described in terms of a pionless effective field theory (\eftnopi), the leading-order parity-violating nucleon-nucleon Lagrangian contains five independent unknown low-energy constants (LECs). We find that imposing the approximate symmetry of QCD that appears when the number of colors $\NC$ becomes large reduces the number of independent LECs to two at leading order in the combined \eftnopi and large-\NC expansions.
We also find a relation between the two isoscalar LECs in the large-$\NC$ limit.
This has important implications for the number of experiments and/or lattice calculations necessary to confirm this description of physics.
In particular, we find that a future measurement of the PV asymmetry in $\vec \gamma d \rightarrow n p$ together with the existing result for PV $\vec{p}p$ scattering would constrain all leading-order (in the combined expansion) LECs. This is considerably improved from the previous understanding of the system.

\end{abstract}

\maketitle

\section{Introduction}

While the attempt to understand parity violation in few-nucleon systems has challenged scientists for decades and generated new experimental techniques and new theoretical
paradigms, only more recently have weak interactions been recognized as a potential tool for probing QCD in few-nucleon systems. Weak interactions are
well understood in isolation on the quark level, but how they are embedded in the nonperturbative environment of the nucleon/nucleus is not.  The hope is that the lever of the weak interaction symmetry can expose QCD behavior.
For reviews of parity violation in nucleon-nucleon interactions, see, e.g., Refs.~\cite{Adelberger:1985ik,RamseyMusolf:2006dz,Haxton:2013aca,Schindler:2013yua}. 

Because QCD is nonperturbative at low energies, attempts to make rigorous predictions in this energy regime typically either involve discretizing QCD and attempting to solve
it numerically on the lattice, or building effective field theories (EFTs) that, while
applicable only in a narrow energy range, include QCD predictions.  EFTs of QCD involve a number of unknown low-energy constants (LECs). The values of these LECs are not fixed by the  symmetries of QCD but must be extracted from 
experiment or calculated on a lattice.  Fortunately, up to  the order we work in the EFT power counting, there are more observables available
(even if not easily measurable) than LECs, which is why the EFT can have predictive power.

An EFT utilizes one or more small
parameters obtained from ratios of disparate scales in the problem, which
are then used as expansion parameters in a perturbative calculation. To
increase calculational power, it is desirable to find as many small parameters
as  possible and to invoke as many symmetries (or near symmetries) of nature
as possible. In addition to small parameters formed from ratios of masses
and momenta, another expansion parameter available in QCD is
 $1/\NC$ in the limit where the number of colors $\NC$ becomes large \cite{tHooft:1973jz,Witten:1979kh}.
Inclusion of this additional expansion typically results in relationships among LECs at a given
order in the EFT power counting, thus reducing the number of LECs that occur at a given order in this combined expansion.  An example is the heavy baryon chiral perturbation theory analysis of the interaction of the octet and decuplet of baryons with the octet of mesons.  
Expansion parameters include $p$ or $m_q$ in the numerator and $\Lambda_\chi$ and $m_B$ in the denominator, with $p$ a typical momentum transfer, $m_q$ a light quark mass, $\Lambda_\chi$ the chiral symmetry breaking scale, and $m_B$ the baryon mass.
Imposing an SU(3) flavor symmetry yields four LECs for describing leading strong interactions.  However, when the approximate symmetry of large $\NC$ is included, only one independent LEC remains in the combined expansion.  Further, measured decay rates indicate that this dual expansion treatment works well \cite{Butler:1992pn}.

At low enough energies, pionless EFT (\eftnopi) is well established as a systematic and rigorous way to determine the restrictions that QCD symmetries place on few-nucleon observables in the parity-conserving (PC) sector. See, e.g., Refs.~\cite{Beane:2000fx,Bedaque:2002mn,Platter:2009gz} for reviews and Ref.~\cite{Rupak:1999rk} for an example of a high-precision calculation.
More recently the methods have been adapted to the parity-violating (PV) sector.  
At energies where the pion mass is not dynamical  (i.e., $E<m_{\pi}^{2}/M_{N}$),  QCD and weak interaction symmetries dictate that five LECs encode all of the physics of parity violation among two nucleons up to and including next-to-leading order (NLO) corrections expected at the 10 percent level \cite{Danilov:1965,Zhu:2004vw,Girlanda:2008ts,Phillips:2008hn}. 

Decades of experimental and theoretical effort have been dedicated to measuring parity violation and understanding its interplay with QCD.
In the EFT framework, this can be viewed as an effort to determine whether the QCD-symmetry-based description is indeed consistent by overconstraining the values of the above LECs.
Given the difficulties in gathering reliable experimental (or lattice QCD\footnote{See Ref.~\cite{Wasem:2011zz} for the first connected diagram calculation of a PV pion-nucleon coupling in lattice QCD.}) information on the LECs, any additional theoretical constraint is valuable.

A general classification of the PV potential in the large-\NC picture was given in Ref.~\cite{Phillips:2014kna}, where they found that there are two leading-order (LO) terms, four terms at NLO, and six terms at next-to-next-to-leading order (NNLO) in the large-\NC counting.
This information was used to analyze PV meson-nucleon couplings that appear in the meson-exchange picture of PV nucleon-nucleon forces \cite{Desplanques:1979hn}.
Here, we consider the implications that a large-\NC analysis has on the LECs of \eftnopi. 
The benefit of the \eftnopi formalism is that it is valid for the experiments we wish to describe, has fewer LECs than theories that contain mesons, and has a well-understood prescription for including higher-order corrections.  The results of this paper continue the efforts of Ref.~\cite{Phillips:2008hn,Schindler:2009wd,Griesshammer:2010nd,Griesshammer:2011md,Schindler:2013yua,Vanasse:2014sva} to express two- and three-body PV processes in a unified description.  
We show that in a combined \eftnopi and large-\NC expansion, the five \eftnopi LECs are not equally important -- only two are of leading order, as expected from the findings of Ref.~\cite{Phillips:2014kna}.
Arriving at this result requires a careful analysis of the Fierz relations that are used when constructing a minimal form of the \eftnopi Lagrangian (for an example see Appendix \ref{App:Fierz}).
One of the two dominant (in the large-\NC expansion) couplings is the isotensor LEC that may be accessible in the reaction $\vec \gamma d \rightarrow np$ at an upgraded High Intensity Gamma-Ray Source (HIGS) at the Triangle Universities Nuclear Laboratory \cite{Ahmed:2013jma,Vanasse:2014sva}. 
In addition, we find that the two isoscalar \eftnopi LECs are equal up to order $1/\NC^2$, i.e., up to corrections expected to be at the 10\% level. 
This presents a significant constraint on the \eftnopi LECs because it matches or surpasses the expected precision of upcoming relevant experiments.

\section{Large-\NC analysis of the parity-violating potential}

The terms contributing to the PV nucleon-nucleon potential up to NNLO in the large-\NC counting were determined in Ref.~\cite{Phillips:2014kna}, which used the definition of the general nucleon-nucleon potential \cite{Kaplan:1996rk}
\begin{equation}
V(\pminus,\pplus) = \langle (\mathbf{p^\prime}_1,\gamma,c), (\mathbf{p^\prime}_2,\delta,d) \vert \hat{H}  \vert (\mathbf{p}_1, \alpha,a) ,(\mathbf{p}_2, \beta, b) \rangle  ,
\end{equation}
where Greek indices denote the spin components and Latin indices the isospin components of the nucleons, respectively.\footnote{This approach does not take into account the effects of virtual baryons, such as the $\Delta$ resonance.  In the large-\NC limit nucleons and the $\Delta$ resonance become degenerate, and as shown in Ref.~\cite{Banerjee:2001js} inclusion of the $\Delta$ is important for the interpretation of the potential in the large-\NC limit to be consistent with meson exchange. Here we project onto the subspace of nucleons, assuming that this does not affect the large-\NC analysis. In a more detailed approach the $\Delta$ resonance would be integrated out \cite{Savage:1996tb,Beane:2002ab}.}
The momenta $\mathbf{p_\pm}$ are given by
\begin{equation}
\mathbf{p_\pm} \equiv \mathbf{p^\prime} \pm \mathbf{p} ,
\end{equation}
where $\mathbf{p^\prime}$ and $\mathbf{p}$ are the outgoing and incoming relative momenta,
\begin{equation}
\mathbf{p^\prime} = \mathbf{p^\prime}_1 - \mathbf{p^\prime}_2, \quad \mathbf{p} = \mathbf{p}_1 - \mathbf{p}_2 .
\end{equation}
As defined, both $\pplus$ and $\pminus$ are P-odd, and $\pplus$ is T-odd, while $\pminus$ is T-even.
The Hamiltonian $\hat{H}$ is the Hartree Hamiltonian, which can be written as \cite{Witten:1979kh,Kaplan:1996rk}
\begin{equation}
\label{eq:Hartree}
\hat{H}=\NC\sum_n\sum_{s,t} v_{stn} \left( \frac{\hat{S}}{\NC} \right)^s  \left( \frac{\hat{I}}{\NC} \right)^t  \left( \frac{\hat{G}}{\NC} \right)^{n-s-t} ,
\end{equation}
where
\begin{align}
\hat{S}^i & = \hat{q}^\dagger \frac{\sigma^i}{2} \hat{q}, & \hat{I}^a & = \hat{q}^\dagger \frac{\tau^a}{2} \hat{q}, & \hat{G}^{ia} & = \hat{q}^\dagger \frac{\sigma^i\tau^a}{4} \hat{q} .
\end{align}
The $\hat{q}$'s denote the doublet of light quarks, which in the large-\NC analysis are colorless bosonic fields. 
The coefficients $v_{stn}$ contain momenta. All vector, spin, and isospin indices (not shown in Eq.~\eqref{eq:Hartree}) are contracted to give the desired symmetry properties.
In the case considered here, we require the resulting potential to be invariant under rotations, P-odd, and T-even.
The large-\NC scaling of the terms in the potential are determined from the matrix elements of the operators $\hat{S}$, $\hat{I}$, $\hat{G}$ , and the identity operator $\mathbb{1}$ between nucleon states, which are given by \cite{Dashen:1994qi,Kaplan:1995yg}
\begin{equation}
\langle  N^\prime \vert  \hat{S}  \vert N  \rangle \sim 
\langle  N^\prime \vert  \hat{I}  \vert N  \rangle \sim 1,  \quad 
\langle  N^\prime \vert  \hat{G} \vert N  \rangle \sim 
\langle  N^\prime \vert \mathbb{1} \vert N  \rangle \sim \NC .
\end{equation}
Products of these operators acting on the same nucleon state can be reduced to single factors of the operators using operator identities and the Wigner-Eckart theorem \cite{Dashen:1994qi}.
In addition, the momenta scale as \cite{Dashen:1994qi,Kaplan:1995yg,Kaplan:1996rk} 
\begin{equation}
\pminus \sim 1\, ,\quad \pplus \sim \NC^{-1} \, .
\end{equation}
The reason for the suppression of $\pplus$ is as follows \cite{Kaplan:1996rk}: it is consistent to interpret the potential derived from the Hartree Hamiltonian in terms of a meson-exchange picture \cite{Banerjee:2001js}.
In the $t$-channel, a factor of $\pplus$ can only appear as a relativistic correction and is therefore always accompanied by an inverse power of the nucleon mass $M_N$. Since $M_N\sim\NC$, $\pplus$ is taken to scale as $\NC^{-1}$. 
In the  $u$-channel the roles of $\pplus$ and $\pminus$ are reversed, but it is common to take this exchange potential into account by considering matrix elements of the potential between states that are properly anti-symmetrized.
A parametrization of the PV potential in terms of one-meson exchanges is given in Ref.~\cite{Desplanques:1979hn}. 
Starting from the relativistic Hamiltonians for PC and PV meson-nucleon interactions given there, a nonrelativistic expansion of the resulting potential satisfies the above counting, i.e., each term proportional to $\pplus$ is suppressed by a factor of $1/M_N$. See Appendix \ref{App:momentum} for details.

Using these ingredients, Ref.~\cite{Phillips:2014kna} determined the operator structures that appear in the PV nucleon-nucleon potential up to NNLO in the large-\NC counting.
The isovector and isotensor interactions include an additional factor of $\sin^2\theta_W \approx 0.23$ \cite{Patrignani:2016xqp}  at the weak matching scale.\footnote{This factor was not included for the isotensor terms in the original manuscript of Ref.~\cite{Phillips:2014kna}.}
While this factor is comparable to $1/\NC$ for the physical case of $\NC=3$, the low energy constants evolve in a nonperturbative (and currently undetermined)  way down into hadronic scales. It is always the case that unknown multipliers can modify the naive ordering of large-\NC terms. In the following we include the nominal factors of $\sin^2\theta_W$ for completeness.
The LO, i.e., $\calO(\NC)$, operator structure is
\begin{equation}
\label{eq:PSSLO}
\pminus \cdot (\vsig_1 \times \vsig_2) \ \vtau_1 \cdot \vtau_2  \  ,  
\end{equation}
while that at $\calO(\NC) \sin^2\theta_W$ is
\begin{equation}
\label{eq:PSSLO_tensor}
\pminus \cdot (\vsig_1 \times \vsig_2) \  \calI_{ab} \tau_1^a \tau_2^b  \  ,
\end{equation}
where $\calI=\text{diag}(1,1,-2)$.
The terms at $\calO(\NC^0) \sin^2\theta_W$ are
\begin{gather}
\label{eq:PSSNLO}
\pplus \cdot (\vsig_1 \tau_1^3-\vsig_2 \tau_2^3) \ , \\
\pminus \cdot (\vsig_1 + \vsig_2) \ (\vtau_1 \times \vtau_2)^3 \ ,  \\
\pminus \cdot (\vsig_1 \times \vsig_2) \ (\vtau_1 + \vtau_2)^3 \ ,  \\
\label{eq:tensor}
\left[ (\pplus\times\pminus)\cdot \vsig_1\, \pminus\cdot\vsig_2 + (\pplus\times\pminus)\cdot \vsig_2\, \pminus\cdot\vsig_1 \right] \ (\vtau_1 \times \vtau_2)^3  \ ,
\end{gather}
while at $\calO(\NC^{-1})$
\begin{gather}
\label{eq:PSSNNLO}
\pminus \cdot (\vsig_1 \times \vsig_2) \  , \\
\label{eq:sup1}
\pplus^2 \ \pminus \cdot (\vsig_1 \times \vsig_2) \ \vtau_1 \cdot \vtau_2 \  , \\
\pplus \cdot (\vsig_1 - \vsig_2) \ , \\
\pplus \cdot (\vsig_1 - \vsig_2) \ \vtau_1 \cdot \vtau_2 \  .
\end{gather}
Two additional isotensor structures contribute at $\calO(\NC^{-1})\sin^2\theta_W$,
\begin{gather}
\label{eq:PSSNNLO_tensor}
\pplus \cdot (\vsig_1 - \vsig_2) \ \calI_{ab} \tau_1^a \tau_2^b  \ ,  \\
\label{eq:sup2}
\pplus^2 \ \pminus \cdot (\vsig_1 \times \vsig_2) \ \calI_{ab} \tau_1^a \tau_2^b  \ .
\end{gather}

In the physical world, $\NC=3$ and the expansion parameter is 1/3. However, in some cases observables or relations between operators are protected from corrections at the next order, which leads to corrections of roughly 10\%. Similarly, the power counting parameter in \eftnopi is  $\frac{p}{\Lambda_{\slashed{\pi}}} \sim 1/3$, with $p$ a typical momentum and $\Lambda_{\slashed{\pi}} \lesssim m_\pi$, but the correction to the leading PV S-P wave mixing  does not occur until two orders higher \cite{Phillips:2008hn}.

\section{Large-\NC analysis of parity-violating low-energy constants in \eftnopi}
\label{sec:LECs}

Our starting point is the LO \eftnopi Lagrangian in the minimal form as given in Ref.~\cite{Girlanda:2008ts},
\begin{equation}
\label{eq:LGir}
\begin{split}
\mathcal{L}_{PV}^\text{min}  = & \ \calG_1 (\N^\dagger \vsig \N \cdot \N^\dagger i \LRd \N -\N^\dagger \N \N^\dagger  \vsig \cdot i \LRd  \N)  \\
& -\tilde \calG_1 \epsilon_{ijk} \N^\dagger \sigma^i \N D^j(\N^\dagger \sigma^k \N) \\
& -\calG_2 \epsilon_{ijk}\left[ \N^\dagger \tau^3\sigma^i \N D^j(\N^\dagger \sigma^k \N) + \N^\dagger \sigma^i \N D^j (\N^\dagger \tau^3\sigma^k \N) \right] \\
& -\tilde \calG_5 \mathcal{I}_{ab}\epsilon_{ijk}\N^\dagger \tau^a\sigma^i \N D^j(\N^\dagger \tau^b \sigma^k \N) \\
& +  \calG_6 \epsilon_{ab3}\vec{D}(\N^\dagger \tau^a \N)\cdot \N^\dagger \tau^b \vsig \N \ ,
\end{split}
\end{equation} 
where 
\begin{equation}
D_\mu \N = \partial_\mu \N +i e \frac{1+\tau^3}{2} A_\mu \N
\end{equation}
is the nucleon covariant derivative and we define $a\, \mathcal{O}\LRd b$ to be\footnote{This definition is not universal. Some authors use the opposite sign convention.}
\begin{equation}
a\, \mathcal{O}\LRd b = a\,\mathcal{O}\vec D b - (\vec D a)\mathcal{O} b \, ,
\end{equation} 
with $\mathcal{O}$ some spin-isospin-operator. 
The $\calG_i$ are related to the $\calC_i$ of Ref.~\cite{Girlanda:2008ts} by $\calG_i = \calC_i/\Lambda_\chi^3$, with $\Lambda_\chi$ the scale of chiral symmetry breaking.
The potential derived from this Lagrangian is
\begin{equation}
\label{eq:VGir}
\begin{split}
V^\text{min}  = &  - \calG_1 \pplus \cdot (\vsig_1-\vsig_2) \\
& - i \tilde\calG_1 \pminus \cdot (\vsig_1\times \vsig_2)  \\
& - i \calG_2 \pminus \cdot (\vsig_1\times \vsig_2) (\tau_1 + \tau_2)^3 \\
& - i \tilde\calG_5 \pminus \cdot (\vsig_1\times \vsig_2) \calI_{ab} \tau_1^a\tau_2^b \\
& +\frac{i}{2} \calG_6 \pminus \cdot (\vsig_1 + \vsig_2) (\tau_1 \times \tau_2)^3 \ .
\end{split}
\end{equation}
Comparing the terms in Eq.~\eqref{eq:VGir} with the terms given in Eqs.~\eqref{eq:PSSLO}-\eqref{eq:sup2},  a naive (and incorrect) assignment of large-\NC scaling to the PV LECs appears to be: 
\begin{gather}
\tilde\calG_5 \sim \NC \sin^2\theta_W, \notag \\
\label{eq:naive} \calG_2\sim \calG_6 \sim \NC^0\sin^2\theta_W ,\\
\calG_1\sim \tilde\calG_1 \sim \NC^{-1} . \notag
\end{gather}
If true, this suggests that in the combined \eftnopi and large-\NC expansion the isoscalar couplings are suppressed compared to the  isotensor and isovector  ones. 
This seems to conflict with the general form of the potential in Eqs.~\eqref{eq:PSSLO}-\eqref{eq:sup2}, which contains an isoscalar part at LO in the large-\NC counting. 

The minimal form of the potential of Eq.~\eqref{eq:VGir} is found by starting with the most general Lorentz-invariant form of the PV Lagrangian with a single derivative and applying Fierz identities as well as a nonrelativistic reduction \cite{Zhu:2004vw,Girlanda:2008ts}.\footnote{In Ref.~\cite{Girlanda:2008ts} the Fierz identities are applied to the relativistic forms of the operators in the basis of Dirac matrices. First performing the nonrelativistic reduction and then applying Fierz identities to the resulting Pauli matrices produces the same set of five nonrelativistic operators.}
Because Fierz identities allow a non-unique set of operators to be expressed as linear combinations of other operators, there exists a freedom to choose which operators to eliminate in favor of others in the ``minimal'' form of the Lagrangian.
From the EFT perspective, all of these choices are equivalent because the Fierz identities only apply to spin-isospin operators, but do not change the number of derivatives or dimensionful quantities such as masses.
On the other hand, the large-\NC counting is related to the spin-isospin structure of the operators, and therefore it is possible to find operators at different orders in 1/\NC that are related by Fierz identities. 
Thus, different minimal forms of the Lagrangian, while equivalent in the EFT counting, can have  different large-\NC scaling.
To find the most conservative estimate of large-\NC behavior, we start from the most general form of the potential, identify the \NC scaling of the operator coefficients, and then apply Fierz identities to reduce the number of operators while maintaining the most dominant \NC scaling for each resulting coefficient.\footnote{We thank L.~Girlanda for discussions on this point, see also \cite{GirlandaCD15}.}

As discussed in Refs.~\cite{Zhu:2004vw,Girlanda:2008ts}, the most general relativistic PV Lagrangian with one derivative contains 12 operators before the application of any Fierz identities.
Two of these operators ($\calO_6$ and $\tilde\calO_6$ in the notation of Ref.~\cite{Zhu:2004vw}) result in the same leading nonrelativistic structure, while two of the other nonrelativistic operators (those proportional to $\tilde C_2$ and $\tilde C_4$ in Ref.~\cite{Zhu:2004vw}) are related by integration by parts.\footnote{This was indicated  in Ref.~\cite{Girlanda:2008ts}.}
Thus, the most general nonrelativistic  Lagrangian contains 10 operators and is given by (cf.~Ref.~\cite{Zhu:2004vw})
\begin{equation}
\label{eq:Lnonmin}
\begin{split}
\mathcal{L}_{PV}^\text{nonmin}  = & \ \calA^+_1 [\N^\dagger \N (\N^\dagger  \vsig \cdot i \LRd  \N) - \N^\dagger \vsig \N \cdot (\N^\dagger i \LRd \N)  ]  \\
& + \calA^-_1 \epsilon_{ijk} \N^\dagger \sigma^i \N D^j(\N^\dagger \sigma^k \N) \\
& + \calA^+_2 [\N^\dagger \N (\N^\dagger  \tau^3 \vsig \cdot i \LRd  \N) - \N^\dagger \tau^3 \vsig \N \cdot (\N^\dagger i \LRd \N)  ]  \\
& + \calA^-_2 \epsilon_{ijk} \N^\dagger \sigma^i \N D^j(\N^\dagger \tau^3 \sigma^k \N) \\
& + \calA^+_3 [ \N^\dagger \tau^a \N (\N^\dagger  \tau^a \vsig \cdot i \LRd  \N) -  \N^\dagger \tau^a \vsig \N \cdot (\N^\dagger \tau^a i \LRd \N) ] \\
& + \calA^-_3 \epsilon_{ijk} \N^\dagger \tau^a \sigma^i\N D^j(\N^\dagger \tau^a \sigma^k \N) \\
& + \calA^+_4 [  \N^\dagger \tau^3\N (\N^\dagger \vsig \cdot i \LRd  \N) - \N^\dagger \vsig \N \cdot (\N^\dagger \tau^3 i \LRd \N)  ]  \\
& + \calA^+_5 \calI_{ab} [  \N^\dagger \tau^a\N (\N^\dagger \tau^b \vsig \cdot i \LRd  \N) - \N^\dagger \tau^a \vsig \N \cdot (\N^\dagger \tau^b i \LRd \N) ]  \\ 
& + \calA^-_5 \mathcal{I}_{ab}\epsilon_{ijk}\N^\dagger \tau^a\sigma^i \N D^j(\N^\dagger \tau^b \sigma^k \N) \\
& + \calA^-_6 \epsilon_{ab3} \N^\dagger \tau^a \N \vec{D}\cdot(\N^\dagger \tau^b \vsig \N) \ ,
\end{split}
\end{equation}
where the $\calA^+_i$ correspond to terms resulting in a $\pplus$ in the potential, while the $\calA^-_i$ terms give a factor of $\pminus$. 
The corresponding potential can be written as (cf.~Ref.~\cite{Zhu:2004vw})
\begin{equation}
\label{eq:Vnonmin}
\begin{split}
V^\text{nonmin}  = &  \   \calA^+_1 \pplus \cdot (\vsig_1-\vsig_2) \\
& +  \calA^-_1 \pminus \cdot i(\vsig_1\times \vsig_2)  \\
& + \calA^+_2 \  \pplus \cdot (\vsig_1 \tau_1^3-\vsig_2 \tau_2^3)   \\
&+  \frac{1}{2}\calA^-_2 \pminus \cdot i(\vsig_1\times \vsig_2) (\tau_1 + \tau_2)^3 \\
& + \calA^+_3 \ \pplus \cdot (\vsig_1 -\vsig_2) \vtau_1\cdot \vtau_2 \\
&+ \calA^-_3 \ \pminus \cdot i (\vsig_1 \times \vsig_2) \vtau_1\cdot \vtau_2  \\
& + \calA^+_4 \ \pplus \cdot (\vsig_1 \tau_2^3 - \vsig_2\tau_1^3)  \\
& + \calA^+_5 \ \pplus \cdot  (\vsig_1-\vsig_2) \calI_{ab} \tau_1^a \tau_2^b \\
& + \calA^-_5 \pminus \cdot i(\vsig_1\times \vsig_2) \calI_{ab} \tau_1^a\tau_2^b \\
& - \frac{1}{2} \calA^-_6 \pminus \cdot (\vsig_1 + \vsig_2) i(\tau_1 \times \tau_2)^3 \ .
\end{split}
\end{equation}
The nonminimal potential $V^\text{nonmin}$ contains all operators  of Eqs.~\eqref{eq:PSSLO}-\eqref{eq:sup2} as identified in Ref.~\cite{Phillips:2014kna} with the exception of Eqs.~\eqref{eq:tensor}, \eqref{eq:sup1}, and \eqref{eq:sup2}.
These operators contain more than one power of momentum and therefore cannot be reproduced by starting from operators with a single derivative.
In addition, the operator structure multiplied by $\calA^+_4$ is of order $\NC^{-2}$ and was not considered in Ref.~\cite{Phillips:2014kna}.
We extract the following large-\NC scaling of the couplings: 
\begin{align}
\label{eq:nonminNC}
\calA^+_1 & \sim \NC^{-1} \ , &  \calA^-_1 & \sim \NC^{-1} \ , \notag \\
\calA^+_2  & \sim  \NC^0\sin^2\theta_W \ , &  \calA^-_2 & \sim \NC^0\sin^2\theta_W \ , \notag \\
\calA^+_3 & \sim \NC^{-1}  \ , & \calA^-_3 & \sim \NC \ , \\
\calA^+_4 & \sim \NC^{-2}  \ ,  \notag \\
\calA^+_5 & \sim \NC^{-1}  \sin^2\theta_W \ , & \calA^-_5 & \sim \NC \sin^2\theta_W \ , \notag \\
& & \calA^-_6 & \sim \NC^0\sin^2\theta_W \ . \notag 
\end{align}
Applying Fierz identities to arrive at the minimal form of Eq.~\eqref{eq:VGir}, the relations between the nonminimal and minimal couplings are
\begin{equation}
\label{eq:nonmin-min}
\begin{split}
\calG_1 & = - \calA^+_1 + \calA^+_3 - 2\calA^-_3  \ , \\
\tilde \calG_1 & = - \calA^-_1 - 2\calA^+_3 + \calA^-_3  \ , \\
\calG_2 & = - \frac{1}{2} \left( \calA^-_2 + \calA^+_2 + \calA^+_4 \right) \ , \\
\tilde\calG_5 & = - \left( \calA^-_5 +  \calA^+_5 \right)  \ , \\
\calG_6 & =  - \calA^-_6 + \calA^+_2 - \calA^+_4  \ .
\end{split}
\end{equation}
A detailed example is worked out in Appendix \ref{App:Fierz}.
The scaling of these couplings is determined by the leading behavior of the LECs $\calA^\pm_i$ of Eq.~\eqref{eq:nonminNC}, which gives
\begin{equation}
\label{eq:minNC}
\begin{split}
\calG_1 &\sim \NC \ , \\
\tilde \calG_1 &\sim \NC \ , \\
\calG_2 &\sim \NC^0\sin^2\theta_W \ , \\
\tilde\calG_5 &\sim \NC \sin^2\theta_W \ , \\
\calG_6 &\sim \NC^0\sin^2\theta_W \ .
\end{split}
\end{equation}
The isotensor coupling is LO in large \NC and the isovector couplings again appear at NLO in large \NC (both multiplied by $\sin^2\theta_W$), as in the naive analysis, see Eq.~\eqref{eq:naive}.
However, the isoscalar couplings are now also counted as LO in large \NC, unlike in the naive analysis.
While it may seem that there are now  two terms at $\calO(\NC)$, the two isoscalar couplings are not independent. 
As seen from Eq.~\eqref{eq:nonmin-min}, the LO contribution to both $\calG_1$ and $\tilde \calG_1$ comes from the coupling $\calA^-_3$, so that up to corrections of order $1/\NC^2$ the relation
\begin{equation}
\label{eq:G1relation}
\calG_1 = - 2 \tilde\calG_1
\end{equation}
holds.
This means that up to corrections on the order of approximately 10\%, two of the five LECs that are independent from the EFT point of view are in fact related to each other. Given the difficulty in determining the LECs from either experiment or lattice QCD and the resulting uncertainties in such an extraction, this result presents a significant simplification.

The physics encoded in the minimal Lagrangian of Ref.~\cite{Girlanda:2008ts} can also be expressed in a  physically more intuitive basis, in which the incoming and outgoing two-nucleon states are in particular partial waves \cite{Phillips:2008hn},
\begin{equation}\label{Lag:PV}
\begin{split}
\mathcal{L}_{PV}=  -  & \left[ \CA \left(\N^T\sigma^2  \vsig \tau^2 \N \right)^\dagger
\cdot  \left(\N^T \sigma^2  \tau^2 i\LRd \N\right) \right. \\
& +\CB \left(\N^T\sigma^2 \tau^2 \vtau \N\right)^\dagger
\left(\N^T\sigma^2  \vsig \cdot \tau^2 \vtau i\LRd  \N\right) \\
& +\CC \ \epsilon_{3ab} \left(\N^T\sigma^2 \tau^2 \tau^a \N\right)^\dagger
\left(\N^T \sigma^2   \vsig \cdot \tau^2 \tau^b \LRd \N\right) \\
& +\CD \ \mathcal{I}_{ab} \left(\N^T\sigma^2 \tau^2 \tau^a \N\right)^\dagger
\left(\N^T \sigma^2  \vsig \cdot \tau^2 \tau^b i \LRd \N\right) \\
& +\left. \CE \ \epsilon_{ijk} \left(\N^T\sigma^2 \sigma^i \tau^2 \N\right)^\dagger
\left(\N^T \sigma^2 \sigma^k \tau^2 \tau^3 \LRd{}^{\!j} \N\right) \right] + \text{H.c.} \, ,
\end{split}
\end{equation}
where $\Delta I$ is the isospin change involved in the process. The relations between the couplings in the two formalisms are \cite{Phillips:2008hn,Vanasse:2011nd}
\begin{equation}
\label{eq:PW-Gir}
\begin{split}
\CA&=\frac{1}{4}(\calG_1-\tilde \calG_1)  \ , \\
\CB&=\frac{1}{4} (\calG_1+\tilde \calG_1) \ , \\
\CC&=\frac{1}{2} \calG_2  \ , \\
\CD&=-\frac{1}{2} \tilde \calG_5 \ , \\
\CE&=\frac{1}{4} \calG_6 \  .
\end{split}
\end{equation}
These relations can be established using Fierz identites, which as discussed earlier can hide the large-\NC scaling of the associated couplings.
However, since we have kept the leading terms for the couplings $\calG_i$, the relations of Eq.~\eqref{eq:PW-Gir} also establish the leading scaling behavior for the partial-wave couplings $\calC^{(X-Y)}$,
\begin{equation}
\label{eq:PWscale}
\begin{split}
\CA & \sim \NC \ , \\
\CB & \sim \NC \ , \\
\CC & \sim \NC^0\sin^2\theta_W  \ , \\
\CD & \sim \NC \sin^2\theta_W \ ,  \\
\CE & \sim \NC^0\sin^2\theta_W  \ .
\end{split}
\end{equation}
As before, the two isoscalar terms are not independent at leading order in the large-\NC counting, but up to $1/\NC^2$ corrections are related by
\begin{equation}
\CA = 3 \, \CB\ .
\end{equation}
Again, the large-\NC analysis shows a relation between two of the five LECs, which is valid at the $\approx 10\%$ level. A determination of one of the isoscalar couplings from either experiment or a future lattice QCD calculation constrains the second isoscalar LEC significantly without need of further experimental or lattice QCD input.

The relations derived above apply to renormalized LECs. As common in \eftnopi, we assume that dimensional regularization in combination with the power divergence subtraction renormalization scheme \cite{Kaplan:1998tg} is applied to obtain any observable.
As shown in Refs.~\cite{Phillips:2008hn,Schindler:2009wd}, the dependence of the PV LECs on the subtraction point $\mu$ is the same as the corresponding S-wave PC LECs, given by \cite{Kaplan:1998tg}
\begin{equation}
\calC^{(\oneS)} \sim \frac{1}{-1/a^{(\oneS)} - \mu} \ ,\quad \calC^{(\threeS)} \sim \frac{1}{-1/a^{(\threeS)} - \mu} \ .
\end{equation}
Here, $a^{(\oneS)}$ and $a^{(\threeS)}$ are the scattering lengths in the singlet and triplet channels, respectively. In the large-\NC limit the LO PC interactions are Wigner-SU(4) symmetric \cite{Wigner:1936dx}, which implies that $\calC^{(\oneS)} =\calC^{(\threeS)} $ in this limit \cite{Kaplan:1995yg}. As a result the $\mu$ dependence of all PC and PV LO LECs becomes identical.
In general, since we are considering the scaling of the LECs with \NC, the relations should hold for any renormalization condition that itself is independent of the number of colors.
However, as pointed out in Ref.~\cite{Kaplan:1995yg}, the approximate Wigner-SU(4) symmetry  is hidden for the physical case of $\NC=3$ if the strong LECs are fit close to threshold, which corresponds to choosing $\mu$ close to zero as the  renormalization condition.
The S-wave scattering lengths are unnaturally large and choosing $\mu=0$ means that the applicability of the \eftnopi expansion is very limited \cite{Kaplan:1998tg}. 
From the EFT point of view it is therefore beneficial to use a renormalization condition in which $\mu$ is of the order of momenta significantly above threshold. 
In these cases, the ratio of PC LECs also gives a much better indication of the approximate Wigner-SU(4) symmetry. 
Because physical observables relevant to hadronic parity violation always involve an interplay of PC and PV interactions, i.e., both PC and PV LECs appear, renormalization conditions in which the consequences of the large-\NC limit are manifest in the PC LECs  should be chosen.


\section{Application to existing measurements}

There has been only one nonzero measurement of a PV two-nucleon observable: the longitudinal asymmetry in $\vec p p$ scattering.\footnote{The NPDGamma \cite{Gericke:2011zz} experiment has finished taking data and a result for the PV angular asymmetry in $\vec{n}p\to d \gamma$ is expected to be announced shortly.} 
The lowest energy result was found in Ref.~\cite{Eversheim:1991tg,EversPrivComm,Haeberli:1995uz}:
\begin{equation}
A^{\vec p p}_L(E=13.6 {\rm \ MeV}) = (-0.93 \pm 0.21) \times 10^{-7} \ .
\end{equation}
 This gives the following constraint on a linear combination of the \eftnopi LECs \cite{Phillips:2008hn}
\begin{equation}\label{muindep}
\frac{4(\CB + \CC + \CD)}{\calC^{(\oneS)}} = (-1.5 \pm 0.3) \times 10^{-10} {\rm \ MeV}^{-1} \ \ , 
\end{equation}
The ratio of the PV and strong LECs is subtraction point independent. Because $\CB$ and $\CD$ dominate in the large-$\NC$ limit, we will neglect $\CC$. 
In particular, one immediate result is that in the large-$\NC$ limit, the longitudinal asymmetry of $\vec p p$ scattering and $\vec nn$ scattering is the same \cite{Phillips:2008hn}.
Imposing the large-$\NC$ relationship between $\CA$ and $\CB$ yields
\begin{equation}
\frac{4(\CA/3 + \CD)}{\calC} = (-1.5 \pm 0.3) \times 10^{-10} {\rm \ MeV}^{-1} \ \ , 
\end{equation}
where $\calC$ denotes the large-\NC limit of $\calC^{(\oneS)}$.
With only one equation and two unknowns, it is not possible to make any statement about the relative size of the two LECs.

However, the reason for expressing $\CB$ in terms of $\CA$ is that an experimental limit exists on  the induced circular polarization in unpolarized neutron capture, $np\to d\vec{\gamma}$ \cite{Knyazkov:1984zz,Knyazkov:1984ke}, which also depends on $\CA$.  Imposing the large-$\NC$ results on the expression for this process \cite{Schindler:2009wd} yields
\begin{equation}
\begin{split}
P_\gamma &= -\frac{16 M_N}{\calC}\frac{1}{\kappa_1 (1-\gamma a^{(\oneS)})}\left(\CA (1- \frac{5}{9} \gamma a^{(\oneS)}) - \frac{2}{3} \gamma a^{(\oneS)} \CD \right) \\
&= (1.8 \pm 1.8) \times 10^{-7}\ \ ,
\end{split}
\end{equation}
where we have also set $\calC^{(\threeS)} \sim \calC$ as dictated by the large $\NC$ limit \cite{Kaplan:1995yg}.
$\gamma$ is the binding momentum of the deuteron, $\kappa_1$ is the isovector anomalous magnetic moment, and $a^{(\oneS)}$ is the singlet channel scattering length. This relationship allows us to say with some confidence that $\CD$ is of the same size as $\CA$.  If it were not, but negligible compared to $\CA$, then the $\vec p p$ scattering measurement would yield
\begin{equation}
\frac{\CA}{\calC} \approx (-1.1 \pm 0.2) \times 10^{-10} {\rm \ MeV}^{-1}\ ,
\end{equation}
and predict a $P_\gamma$ larger than its present bound. This is also an indication that the factor of $\sin^2\theta_W$ that multiplies the isotensor interaction at the weak matching scale is not significant at hadronic scales.


\section{Conclusions}
\label{sec:Con}

We have shown that a large-$\NC$ symmetry imposed on the \eftnopi description of low-energy parity violation in two-nucleon systems through NLO in \eftnopi power counting reduces the number of independent LECs from five to two at LO in  the combined expansion.  
This is in agreement with the general large-\NC analysis of PV nucleon-nucleon forces of Ref.~\cite{Phillips:2014kna}. However, we find the result expressed in the \eftnopi formalism to be more useful for interpreting experimental measurements in few-nucleon systems, both because there are fewer LECs and because analyzing higher-order corrections is straightforward. 
Further, the result in the \eftnopi formalism is not obvious from a naive application of the large-\NC counting rules to the minimal form of the \eftnopi potential. The reason is that the Fierz identities used in reducing the most general Lagrangian to its minimal form can hide the order in large-\NC counting at which a given term first contributes.
We have also found a relation between the two isoscalar LECs $\CA$ and $\CB$ that is expected to hold up to corrections of order $1/\NC^2$, i.e., on the order of 10\%.

This could have important implications for how quickly parity violation in two- and few-nucleon systems can be understood.  In particular, the existing measurement of the longitudinal asymmetry in $\vec p p$ scattering along with the limit of the $np\rightarrow d\vec{\gamma}$ PV asymmetry suggests that the two PV LECs that are of LO in the combined \eftnopi and large-\NC expansion, $\CA$ and $\CD$, are in fact of the same size.
That is, current experiments support the leading-order analysis.
One linear combination of these two LECs can be determined from the existing experimental result on $\vec{p}p$ scattering.
If the result that these two LECs are of the same size holds when higher-order \eftnopi and large-$\NC$ corrections are included, it is a motivation to measure $\CD$ at the potential HIGS2 facility \cite{Ahmed:2013jma} and gives further motivation to perform a lattice calculation of $\CD$ \cite{FunSymLattPos}.
Because of the large-\NC relation between $\CA$ and $\CB$, knowledge of $\CD$ would then constrain both isoscalar LECs.

\appendix
\section{Example of the application of Fierz identities}
\label{App:Fierz}

As an example of how to reduce the nonminimal \eftnopi Lagrangian to its minimal form, consider the term proportional to $\calA^-_3$,  
\begin{equation}
\epsilon_{ijk} \N^\dagger \tau^A \sigma^i \N D^j  (\N^\dagger \tau^A \sigma^k \N) \,,
\end{equation} 
which in components reads
\begin{equation}
(\tau^A_{ab} \tau^A_{cd}) (\epsilon_{ijk}\sigma^i_{\alpha \beta} \sigma^k_{\gamma\delta}) \N_{\alpha,a}^\dagger \N_{\beta,b} D^j(\N_{\gamma,c}^\dagger \N_{\delta,d}).
\end{equation} 
Applying the Fierz identity
\begin{equation}
\tau^A_{ab} \tau^A_{cd} = 2\delta_{ad} \delta_{cb} -\delta_{ab} \delta_{cd},
\end{equation}
the second term immediately gives 
\begin{equation}
-\epsilon_{ijk} \N^\dagger\sigma^i \N D^j \cdot (\N^\dagger \sigma^k \N)  \  ,
\end{equation} 
i.e., the operator structure proportional to $\calA^-_1$.
For the remaining term, the Fierz identity in spin space
\begin{equation}
\sigma^i_{\alpha \beta} \sigma^k_{\gamma\delta} = \frac{1}{2} \left[ \sigma^i_{\alpha \delta} \sigma^k_{\gamma\beta} + \sigma^k_{\alpha \delta} \sigma^i_{\gamma\beta} 
+\delta^{ik} (\delta_{\alpha\delta} \delta_{\gamma\beta} - \sigma^l_{\alpha \delta} \sigma^l_{\gamma\beta}) 
+ i {\epsilon^{ik}}_l ( \sigma^l_{\alpha \delta} \delta_{\gamma\beta} -  \delta_{\alpha \delta} \sigma^l_{\gamma\beta})\right]
\end{equation}
can be applied.
In addition, using integration by parts and dropping the term proportional to a total derivative, the combination of nucleon fields can be rewritten as
\begin{equation}
\begin{split}
N_{\alpha,a}^\dagger \N_{\beta,b} D^j(\N_{\gamma,c}^\dagger \N_{\delta,d}) & = \frac{1}{2} \left[\N_{\alpha,a}^\dagger \N_{\beta,b} D^j(\N_{\gamma,c}^\dagger \N_{\delta,d}) - D^j (\N_{\alpha,a}^\dagger \N_{\beta,b}) (\N_{\gamma,c}^\dagger \N_{\delta,d}) \right]\\
& =  \frac{1}{2} \left[ \N_{\alpha,a}^\dagger  \N_{\delta,d} ( \N_{\gamma,c}^\dagger \LRd \N_{\beta,b}) - ( \N_{\alpha,a}^\dagger \LRd \N_{\delta,d} ) \N_{\gamma,c}^\dagger \N_{\beta,b} ) \right].
\end{split}
\end{equation}
The resulting term has the same form as the operator structure proportional to $\calA^+_1$. 
Therefore, the term proportional to $\calA^-_3$ can be removed by a redefinition of  $\calA^+_1$ and $\calA^-_1$.


\section{Momentum dependence in the PV meson-exchange potential }
\label{App:momentum}

As an example of how the suppression of terms proportional to $\pplus$ with a factor of $1/M_N$ arises in the meson-exchange picture, we consider the PV potential as given in Ref.~\cite{Desplanques:1979hn}, which parameterizes the PV interactions in terms of $\pi$, $\rho$, and $\omega$ exchange.
Here we further restrict the discussion to $\omega$ exchange, but analogous arguments apply to $\pi$ and $\rho$ exchanges. The PC Hamiltonian is given by \cite{Desplanques:1979hn}
\begin{equation}
\mathcal{H}_{PC}^\omega = g_\omega\bar{\Psi}\left( \gamma^\mu + i \frac{\chi_S}{2\Lambda_\chi} \sigma^{\mu\nu}k_\nu \right) \omega_\mu \Psi \, ,
\end{equation} 
where compared to Ref.~\cite{Desplanques:1979hn} we have replaced a factor of $M_N\sim \NC$ by $\Lambda_\chi\sim 1$ in the denominator (see the discussion in Ref.~\cite{Phillips:2014kna}) because the origin of this scale is not dynamical and should not introduce spurious \NC dependence. 
The PV Hamiltonian reads
\begin{equation}
\mathcal{H}_{PV}^\omega = \bar{\Psi} \left(h_\omega^0 \omega_\mu + h_\omega^1 \tau^3 \omega_\mu  \right) \gamma^\mu\gamma_5 \Psi \, .
\end{equation}
Under a nonrelativistic expansion, the terms proportional to $g_\omega$ are LO for $\mu=0$ and suppressed by $1/M_N$ for $\mu=i$, while terms proportional to $g_\omega\chi_S$ scale as $1/M_N$ for $(\mu\nu) = (0i)$ and are LO for $(\mu\nu)=(ij)$.
Analogously, the PV terms scale as $1/M_N$ for $\mu=0$ and are LO for $\mu=i$.
Combining these expressions, the terms proportional to $\pplus$ arise in combination with $g_\omega h^{0,1}_\omega$ and are suppressed by $1/M_N$, while terms proportional to $\pminus$ come with $g_\omega\chi_S h^{0,1}_\omega$ and are LO in the nonrelativistic expansion.
Analogous relations hold for $\pi$ and $\rho$ exchanges.

\begin{acknowledgments}
We would like to thank L.~Girlanda, D.~R.~Phillips, and C.~Schat  for useful discussions.
This material is based upon work supported by the U.S. Department of
Energy, Office of Science, Office of Nuclear Physics, under Award Number DE-SC0010300 (MRS), Award Number DE-FG02-05ER41368 (RPS and JV), and Award Number DE-FG02-93ER40756 (JV).
\end{acknowledgments}

\end{document}